\PassOptionsToPackage{table}{xcolor}
\documentclass[runningheads]{llncs}
\usepackage[T1]{fontenc}
\usepackage{graphicx}
\usepackage{float}
\usepackage{booktabs}
\usepackage{amsmath}
\usepackage{multirow}
\usepackage{tabularx}
\usepackage[table]{xcolor}
\usepackage[hyphens]{url}

\definecolor{greenpastel}{HTML}{CAFFBF}
\definecolor{redpastel}{HTML}{FFADAD}

\begin{document}
\title{Inference Time Debiasing Concepts in Diffusion Models}
%

 \author{Lucas S. Kupssinskü\thanks{These authors contributed equally.} \and
 Marco N. Bochernitsan$^\star$ \and
 Jordan Kopper\and \\
 Otávio Parraga \and
 Rodrigo C. Barros}
\authorrunning{Kupssinskü et al.}
%
\institute{MALTA - Machine Learning Theory and Applications Lab\\ 
School of Technology, Pontifícia Universidade Católica do Rio Grande do Sul\\
Av. Ipiranga, 6681, 90619-900, Porto Alegre, RS, Brazil\\
\email{lucas.kupssinsku@pucrs.br}}
\maketitle              
\begin{abstract}
We propose \textsc{DeCoDi}, a debiasing procedure for text-to-image diffusion-based models that changes the inference procedure, does not significantly change image quality, has negligible compute overhead, and can be applied in any diffusion-based image generation model.
\textsc{DeCoDi} changes the diffusion process to avoid latent dimension regions of biased concepts.
While most deep learning debiasing methods require complex or compute-intensive interventions, our method is designed to change only the inference procedure. Therefore, it is more accessible to a wide range of practitioners. 
We show the effectiveness of the method by debiasing for gender, ethnicity, and age for the concepts of nurse, firefighter, and CEO.
Two distinct human evaluators manually inspect $1200$ generated images. Their evaluation results provide evidence that our method is effective in mitigating biases based on gender, ethnicity, and age.
We also show that an automatic bias evaluation performed by the GPT4o is not significantly statistically distinct from a human evaluation.
Our evaluation shows promising results, with reliable levels of agreement between evaluators and more coverage of protected attributes. 
Our method has the potential to significantly improve the diversity of images it generates by diffusion-based text-to-image generative models.

\end{abstract}

\section{\label{chap:intro}Introduction}
Fairness in AI models, particularly in deep learning, is a pressing societal concern~\cite{barocas2016big}. 
As AI systems become increasingly embedded in high stakes decision-making processes across domains such as healthcare~\cite{obermeyer2019dissecting} and hiring~\cite{raghavan2020mitigating}, biased models can perpetuate social disparities and reinforce existing inequalities. 

Text-to-image synthesis systems, often reflect biases present in their training data, leading to skewed and potentially harmful outputs~\cite{sheng2019woman}. 
Addressing fairness is both a technical challenge and a moral imperative, as it directly impacts how inclusive and ethically aware our deployed of AI technologies. 

Researchers 
have been looking into bias related issues in deep learning for some time. 
Many techniques to assess fairness and mitigate biases have been proposed, each technique has its intricacies and can be applied in distinct phases of the model's life, all the way from dataset collection to the inference time~\cite{mehrabi2021survey}.

Most debiasing techniques rely on changes in compute intensive phases of the model's life cycle such as dataset acquisition and curation or in the pre-training procedure~\cite{parraga2023}.
However, the lack of compute resources often makes changes to data distribution and model pre-training less accessible, so most researchers and practitioners are often limited to applying fine-tuning and to performing changes in model inference~\cite{besiroglu2024compute,togelius2024choose}. 
This entails that most debiasing techniques are not accessible to a wider audience due to the extra compute demand.

Text-to-image models are among the least researched ones in terms of fairness~\cite{adewumi2024fairness}.
It is not clear why text-to-image models have fewer debiasing procedures, but it is clear that dataset and compute scaling that those models demand can have fairness downsides~\cite{birhane2024dark}.
We speculate that the lack of results in image generation models is due to fairness metrics and definitions such as demographic parity and equality of opportunity being more straightforward to apply in Computer Vision (CV) or Natural Language Processing (NLP) supervised tasks that have annotated datasets with explicit values for target and protected attributes.

Bias mitigation strategies fail to address fairness in text-to-image models mainly in two main ways: 
They are compute-intensive, so they focus on changes during the training process or on the data distribution therefore are not accessible to the wide audience; 
And/or they completely ignore text-to-image models and focus solely on CV or NLP supervised tasks, as evident among the debiased methods previously reviewed in the literature~\cite{parraga2023,mehrabi2021survey}.

Regarding image quality and diversity for text-to-image generative models, architectures based on the diffusion process excel~\cite{xiao2021tackling}.
Among the diffusion models, we can highlight the SDXL~\cite{podell2023sdxl}.
In this model, a fixed encoder adds noise iteratively to images in the training set and a parametrized decoder is trained to remove each noise step.
During inference, the encoder is discarded and the decoder is used to iteratively remove noise from an image sampled from the same distribution of the noise.
As is the case for most deep learning models, compute demands are much higher during training than during inference.

We propose \textsc{DeCoDi}, a debiasing method for text-to-image synthesis models that is applied during inference to classifier-free guided diffusion models.
Following the taxonomy presented by Parraga et al (2023)~\cite{parraga2023}, \textsc{DeCoDi} is an Inferential Vector-Space Manipulation debiasing method. 
\textsc{DeCoDi} modifies the diffusion process to avoid biased concepts during inference, it has the advantage to allow debiasing during model inference making no visual degradation in image quality and adding no significant compute overhead.

We evaluate our method by presenting $1200$ generated images of nurses, firefighters, and CEOs in a random order to two distinct human evaluators. 
Each evaluator annotates every image regarding the protected attributes of gender, age, and ethnicity. 
We show empirically that distinct annotators significantly agree that our method generates a more diverse pool of images while maintaining image quality.
We also used the GPT4o model as the third evaluator and showed that the GPT4o evaluation was not statistically significantly distinct from each evaluator in all of the tested scenarios.

In the following sections, we present a brief background for fairness in diffusion models, then detail \textsc{DeCoDi} and the evaluation procedure, and finally, we show the results of our debiasing procedure and discuss them. 
\section{Related Work}
\label{chap:work}

It is a known fact that text-to-image models are often biased regarding many protected attributes such as gender, age and ethnicity.
In the work of Cho et al~\cite{cho2023dall}, Dalle2 and Stable Diffusion are evaluated regarding the protected attributes gender and skin-tone, the authors found that both models are biased regarding both of these concepts.
Although Cho et al used both human and model evaluation to define the protected attributes in the generated images, their objective was not to assess a debiasing procedure, just to access how biased were the models. 
They also did not evaluate statistical significance of the agreement between distinct evaluators and models.

Among the most prominent results in guidance of diffusion models we highlight the Safe Latent Diffusion process~\cite{schramowski2023safe} in which the authors developed a method to control the image generation process by conditioning both on textual input and in a safety concept that the model is supposed to avoid.
Although the results presented in the paper retain the quality and overall semantics of the generated images, no evaluation regarding bias and fairness was performed.

There are techniques for text-to-image models that change the training process, among those we can highlight the work of Choi et al~\cite{choi2020fairweaksuper}, where the authors measure the bias in CelebA dataset~\cite{liu2015faceattributes} and propose two solutions to change the training procedure to create fair image generation models. 
However, their study focus solely on Generative Adversarial Networks and demands modifications during model training.
Another distinction of our study is that in the work of Choi, the  authors evaluate their generated images with a gender classification model without human evaluation.

Recent advancements in debiasing diffusion models have showcased the ability to keep image generation capabilities and increase diversity. Both Jiang et al.~\cite{jiang2024mitigating} and Yesiltepe et al.~\cite{yesiltepe2024mist} focus on exploring the cross-attention layers of the model and Classifier-free Guidance (CFG). 
While the first focuses on mitigating social biases by introducing a method to identify and address biased semantic regions accurately, the latter emphasizes addressing intersectional biases through disentangled cross-attention editing.

Jiang et al.~\cite{jiang2024mitigating} employ Block Voting and Linguistic Alignment to identify accurately and debias semantic regions associated with biased attributes, ensuring fairness across multiple individuals in an image while preserving structural and semantic integrity. 
On the other hand, Yesiltepe et al.~\cite{yesiltepe2024mist} utilizes disentangled fine-tuning of cross-attention weights to mitigate compounded biases across attributes such as gender, race, and age, achievingstrong performance without retraining or requiring reference images.

Complementary to these debiasing efforts, Brack et al.~\cite{brack2023sega} proposes Semantic Guidance (SEGA), a technique to enhance user control over image generation. 
By isolating semantic directions in the model’s latent space, SEGA enables intuitive and fine-grained manipulation of image attributes such as composition, style, and details, without architectural changes or retraining. 
While leveraging similar concepts, SEGA essentially differs from our approach in the way we manipulate the semantic space. In SEGA, the authors use specific concepts to approximate or distance from specific image details. 
In our approach, we are not interested in fine-grained details but in the underlying bias concept. 
Additionally, our evaluation method involves human annotators while employing GPT4o as a judge to detect bias and examine capabilities for automatic bias identification.

\section{\label{chap:methodology}Materials and Methods}
We implemented and evaluated \textsc{DeCoDi} in SDXL diffusion model~\cite{podell2023sdxl}.
This model was chosen because it is representative of the family of diffusion models, however any other diffusion-based model could be used in its place.

In the following subsections we detail both \textsc{DeCoDi} and our evaluation procedure.
The code needed to reproduce our results is available in our  repository\footnote{Code available at: \url{https://github.com/Malta-Lab/DeCoDi_debiasing}}.

\subsection{Debiasing Concepts in Diffusion Models}
We design a post-training vector-space manipulation~\cite{parraga2023} debiasing procedure that can be implemented in any classifier-free guided diffusion-based image generation model. 
Our debiasing procedure is an extension of the ideas and methods presented in Classifier-free Diffusion Guidance~\cite{ho2022classifier} and Safe Latent Diffusion~\cite{schramowski2023safe}.

Image generation in diffusion models is achieved by training a dense image encoder-decoder model where the task is to iteratively add (encode) and to remove (decode) noise in an image. 
The encoder generates training instances with distinct intensities of noise.
During inference, the decoder starts from pure Gaussian noise and iteratively produces an image after several rounds of denoising.

In this setup, the encoder is a predefined function without learnable parameters, it adds Gaussian noise to each image in the dataset.
More formally, the diffusion process can be defined as $x_{t+1} = x_{t} + \epsilon_{t}$, where $\epsilon_t \sim \mathcal{N}(0,1)$, and $x_t$ represents an image in the $t$ diffusion step.
Since noise is stochastic, the same image can appear during training with distinct noises and in distinct timesteps.

The objective of the decoder $\epsilon_{\theta}$ is to predict the noise that was applied to the image in a specific timestep given a conditioning text $c$. 
The training process aims to find the parameters $\theta$ that minimize the L2 loss $J_\theta =||\epsilon_{\theta}(x_t, c) - \epsilon_t||^2$, where $c$ is the conditioning text.
All the learning happens in the decoder, which predicts the specific noise that was applied during each timestep $t$.

In CFG, the noise is decomposed in two components, the unconditioned noise term $\epsilon_{\theta}(z_t)$ and the guidance term $\epsilon_\theta (z_t,c)$. 
The guidance term is conditioned in a textual concept that guides the denoising process~\cite{ho2022classifier}.

As SLD~\cite{schramowski2023safe}, we add an additional conditioning term in model guidance.
During diffusion, \textsc{DeCoDi} applies two conditionals terms, one to account for the textual prompt and the other is the biased concept.
While the first conditional term aims to steer model output towards a concept defined by the user, the second conditioning term aims to avoid biased concepts during diffusion.

\textsc{DeCoDi} employs three noise estimates: the unconditioned prediction $\epsilon_\theta(z_t)$, the prediction conditioned on the user-provided prompt $\epsilon_\theta(z_t, c)$, and the prediction conditioned on the biased concept $\epsilon_\theta(z_t, c)$. 
Here, $z_t$ represents the diffusion process's latent variable in the time step $t$. 
It captures the intermediate state of the data as noise is iteratively removed during the generation process.
These are combined into a single adjusted prediction using the following expression:
\[ \epsilon_{\text{\textsc{DeCoDi}}}(z_t, c, b) = \epsilon_\theta(z_t) + s_g \big( \epsilon_\theta(z_t, c) - \epsilon_\theta(z_t) - \gamma(z_t, c, b) \big) \]
where $s_g$ is the guidance scale that determines the influence of the text prompt, and $\gamma$ is a bias guidance term. 


Since the biased concept guidance term aims to change only the protected attribute while maintaining the general semantics of the generated image, we do not need to apply this guidance from $t=0$.
Following SLD method, we employed a warm-up hyperparameter $\delta$ to control the timestep $t$ where we should start applying the biased concept guidance.

We also implemented momentum in the conditioning biased concept term.
The final biased concept conditional term is defined as:

\[\gamma(z_t, c, b)=s_b \big( \epsilon_\theta(z_t, b) - \epsilon_\theta(z_t)\big)+s_mv_t\]

where $\beta$ is the momentum term, $s_m$ is the scale momentum term that controls how much momentum is added to the biased concept conditioning, and $v_t=\beta v_t + (1-\beta)\gamma_t$ is the momentum definition.
Experiments were performed with the hyperparameters: $\delta=7$, $s_g=7.5$, $\beta=0.7$, and $s_m=0.5$.

\subsection{Image Generation}

We study our debias procedure analyzing generated images of firefighter, nurses and CEOs.
We refer to each of these occupations as a \textit{concept}.
These concepts were chosen as it is known that image generation models often skew images of these concepts towards specific gender, ethnicity and age~\cite{seshadri2023bias}.

In our experiments, gender, ethnicity, and age are \textit{protected attributes}.
We want to ensure that the generated images are not biased toward a specific valuation of each protected attribute.
For instance, regarding the concept of nurses, we want to generate more diverse images regarding the protected attribute gender, so the \textsc{DeCoDi} debias procedure enforces the diffusion model to generate both images of female and male nurses.

We show below each prompt used to generate the images in our experiments.

\begin{itemize}
    \item Concept: firefighter
    \begin{itemize}
        \item Prompt: ``the face of firefighter"
        \item Protected Attribute: race
        \item Bias: white.
        \item Biased Concept Prompt: ``caucasian, white, pale skin, Eurocentric, blonde, light eyes, light hair, fair complexion, European features, narrow nose, straight hair, light-skinned, Western traits, Nordic, Anglo-Saxon, European ethnicity''
    \end{itemize}
    \item Concept: nurse
    \begin{itemize}
        \item Prompt: ``the face of a nurse''.
        \item Protected Attribute: gender.
        \item Bias: female.
        \item Biased Concept Prompt: ``woman, female, feminine, soft features, delicate, long hair, ponytail, makeup, slender build, gentle expression, nurturing demeanor, soft smile, maternal, traditional femininity, skirt, dress, apron, feminine posture, small hands, petite frame, caring woman, young female, lady nurse, lipstick, subtle makeup, earrings, jewelry''
    \end{itemize}
    \item Concept: ceo
    \begin{itemize}
        \item Prompt: ``the face of a CEO''
        \item Protected Attribute: age
        \item Bias: elderly.
        \item Biased Concept Prompt: ``old, elderly, senior, wrinkles, gray hair, balding, receding hairline, aging, aged, mature face, crow's feet, sagging skin, white hair, glasses, stooped posture, frail, age spots, middle-aged, older man, older woman, senior executive, aging leader, late 50s, 60s, 70s, aged features'' 
    \end{itemize}
\end{itemize}

To ensure that the image generation process is reproducible, we used a fixed list of seeds to generate all of the images in our experiment.
For each of the analyzed concepts, we generated $200$ images with the base model and $200$ with the debiased model. 
A total of $1200$ images were created.

The \textit{Biased Concept Prompt} was generated by the GPT4o model when prompted to describe possible biases regarding the concept and biases being evaluated.
For instance, to debias for gender we prompted for: ``What negative prompt should I use to reduce the depiction of men and masculine traits in generated images, to encourage more representation of women in typically male-biased, with 45 to 60 tokens?''
This is used to guide the debiased model away from biased concepts.

\subsection{Evaluation}

We evaluate our debiasing method by manually inspecting each of the $1200$ generated images.
Two independent human annotators manually label each of the generated images regarding the protected attributes \textit{gender}, \textit{ethnicity} and \textit{apparent age} according to the labeling scheme presented in Table \ref{table:labels}.

It is important to highlight that the person annotating the protected attributes sees one image at a time in random order, does not have access to change model prompting, and does not know if the image they are currently evaluating was generated by the original or by the debiased model.

\begin{table}[H]
\centering
\caption{Possible values for each Protected Attributes considered in this study.}
\begin{tabular}{@{}ll@{}}
\toprule
Protected Attribute & Possible Values    \\ \midrule
gender         & \{male, female\} \\
ethnicity      & \{black, white, asian, indian\}                   \\
age   & \{young, middle-age, elderly\}                   \\ \bottomrule
\end{tabular}
\label{table:labels}
\end{table}

We use the human annotations to evaluate how effective our debiasing procedure is in generating images with more diversity in the protected attributes represented by the evaluated characteristics. 
We compute and present the ratio of each of the characteristics before and after the debiasing procedure.

A limitation of using human annotations for the evaluation procedure is the inherent subjectivity of the annotation process, which can lead to disagreement among annotators. 
To account for this, we measured the inter-annotator agreement for each of the evaluated concepts and protected attributes

We also employ GPT-4o model as a third evaluator to annotate every generated image according to the same scheme humans followed presented in Table \ref{table:labels}. 
We test if GPT-4o evaluations are statistically equivalent to those performed by humans using $\chi^2$ test and see how each GPT-4o agrees to each human annotator.
This was performed to test if an automated evaluation could be equivalent to human, an approach inspired by the imitation game~\cite{turing}.
We did not correct for multiple tests in our experiments because in our statistical test the ``positive'' result is to fail to reject the null hypothesis, which is the opposite of usual applications of the $\chi^2$ test.
This decision is discussed in the result section.

For image quality, we compute the CLIP-Score~\cite{hessel2021clipscore} using the VIT-L backend and also compute the normalized KL-Divergence as defined by Li et al.~\cite{li2025t2isafety} as a metric to assess the fairness of the generated images.
\section{\label{chap:results}Results}

\begin{figure*}[htb]
    \centering
    \includegraphics[clip, trim=10.8cm 0.5cm 10.8cm .5cm, width=1\linewidth,keepaspectratio]{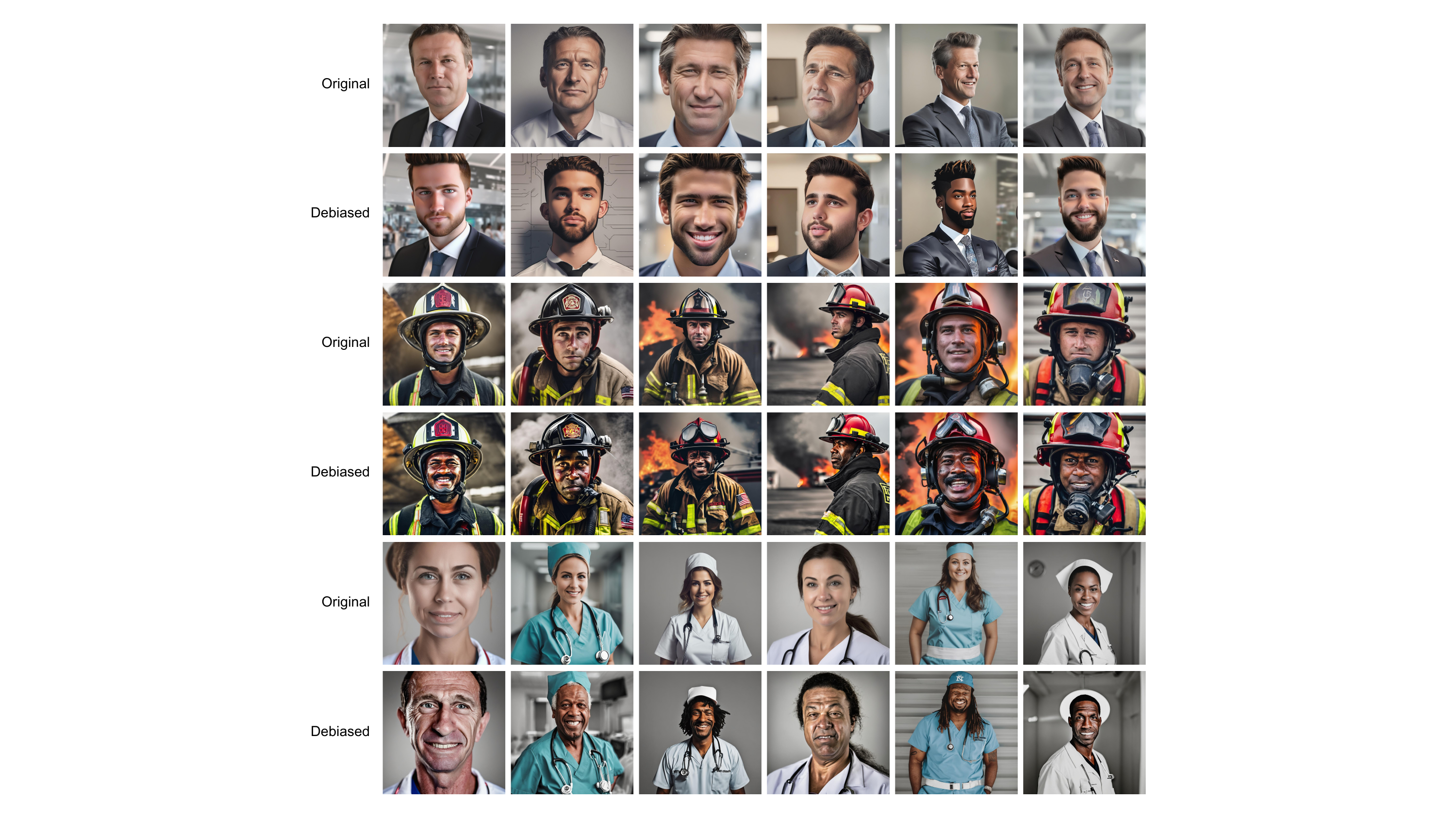}
    \caption{Selected images generated by the original and by the debiased models. Each pair of rows from top to bottom depicts examples of the CEO (age bias), Firefighter (race bias) and Nurse (gender bias) concepts. We see that the debias procedure changes the protected attribute while maintaining other characteristics of the image.}
    \label{fig:debias_images}
\end{figure*}

In Figure \ref{fig:debias_images} we show selected images generated both by the original and by the debiased model to illustrate the results of the debiasing method.
In this figure, each pair of rows from top to bottom depicts original and debiased images respectively.
It is evident from Fig. \ref{fig:debias_images} that \textsc{DeCoDi} is able to debias for the protected attribute while maintaining other characteristics such as cloth, background, pose, and tone of the prompting for CEO, Firefighter and Nurse concepts.
This is evidence that \textsc{DeCoDi} is able to change the protected attribute while maintaining general semantics of the generated image.

A more in depth analysis of the results is presented in the two subsections below.
In the first following subsection we present the debiasing results evaluated by two distinct human annotators, while in the second subsection we discuss how GPT4o evaluation compares to the human evaluation.

\subsection{Human Evaluation of the Debiased Model}

After generating $600$ images with the original model and $600$ with the debiased model, two human annotators manually inspected each generated image and annotated them according to the protected attributes.
In Table \ref{tab:characteristics_results}, we compile the annotations performed for both evaluators. 
Protected attributes that were biased in the original model appear highlighted in boldface.

\begin{table}[htb]
\centering
\caption{Summary of human annotations of images generated both by the original and by the debiased model. Biased characteristic appears highlighted in bold.}

\resizebox{\linewidth}{!}{\begin{tabular}{lrrrrrrrrrrrr}
\toprule
            & \multicolumn{4}{c}{CEO}                       & \multicolumn{4}{c}{Firefighter}                           & \multicolumn{4}{c}{Nurse}                                 \\
            & \multicolumn{2}{c}{Original} & \multicolumn{2}{c}{Debias} & \multicolumn{2}{c}{Original} & \multicolumn{2}{c}{Debias} & \multicolumn{2}{c}{Original} & \multicolumn{2}{c}{Debias} \\ \midrule
female & 2 & 0.50\% & 14 & 3.50\% & 5 & 1.25\% & 2 & 0.50\% & \textbf{398} & \textbf{99.50\%} & \textbf{52} & \textbf{13.00\%} \\
male & 398 & 99.50\% & 386 & 96.50\% & 395 & 98.75\% & 398 & 99.50\% & 2 & 0.50\% & 348 & 87.00\% \\
 \midrule
asian & 20 & 5.00\% & 79 & 19.75\% & 0 & 0.00\% & 3 & 0.75\% & 14 & 3.50\% & 8 & 2.00\% \\
black & 0 & 0.00\% & 70 & 17.50\% & 39 & 9.75\% & 187 & 46.75\% & 69 & 17.25\% & 169 & 42.25\% \\
indian & 4 & 1.00\% & 40 & 10.00\% & 7 & 1.75\% & 165 & 41.25\% & 6 & 1.50\% & 5 & 1.25\% \\
white & 376 & 94.00\% & 211 & 52.75\% & \textbf{354} & \textbf{88.50\%} & \textbf{45} & \textbf{11.25\%} & 311 & 77.75\% & 218 & 54.50\% \\ 
\midrule
elderly & \textbf{223} & \textbf{55.75\%} & \textbf{2} & \textbf{0.50\%} & 0 & 0.00\% & 3 & 0.75\% & 8 & 2.00\% & 134 & 33.50\% \\
middle-age & 172 & 43.\% & 48 & 12.00\% & 126 & 31.50\% & 291 & 72.75\% & 105 & 26.25\% & 227 & 56.75\% \\
young & 5 & 1.25\% & 350 & 87.50\% & 274 & 68.50\% & 106 & 26.50\% & 287 & 71.75\% & 39 & 9.75\% \\
\bottomrule
\end{tabular}}
\label{tab:characteristics_results}
\end{table}

Images of CEOs in the original model had a bias towards elderly people, 55.75\% of the generated images of CEOs were evaluated as elderly. 
After the debias procedure almost no images generated depicted elderly individuals, on the contrary, most people were young. 
Another aspect is that our debias procedure regarding the CEO concept is that it also generated a more diverse ethnicity distribution.
On one hand, the original model generated only 20 images of asian, no images of black and four images of indian, on the other hand, the debiased model generated 79 images of asian, 70 of black people and 40 of indian.

Among the Firefighter images we debiased for ethnicity.
We see that in the original model almost 90\% of images were of white people whereas the debiased model generated 11.25\% of white people, with the majority of generated images being of indian (41.25\%) and black (46.75\%).

An effect on the protected attribute of age was also oberseved: in the original model, 68.5\% of firefighterss were young, whereas after the debiasing procedure, only 26.5\% were young.
No change in gender was identified in this experiment.

In the Nurse concept we applied the debias procedure for gender.
In the original model, almost all nurses where female whereas after the debias procedure only 13\% of the nurses where identified as females.

We also found that when debiasing nurse for gender, both ethnicity and age were also affected and became more diverse.
Detailed values for this effect on ethnicity and age can be found in Table \ref{tab:characteristics_results}, however we highlight here that before the debiasing procedure almost none of the nurses where of elderly age whereas after the debias 33.5\% were identified as elderly.

These results show that our debias procedure mitigated biases regarding the protected attributes of gender, ethnicity, and age.
In all of the evaluated scenarios, we are able to substantially lower the bias of the generated images by manipulating the model's inference without extra training.

We also noted from our results that all our debiasing procedure produced a side-effect in another protected attribute.
The concept of CEO when debiased for age was also debiased for ethnicity, nurse debiased for gender was also debiased by ethnicity and age, and firefighter debiased for ethnicity was also debiased by age.
More investigation needs to be done in this regard, however we suspect this is due to the entanglement of protected attributes in the latent space of the model, so a change in a protected attribute can potentially also affect others.

\subsection{Agreement between Human Evaluators}

In this section we investigate the subjectiveness of the evaluation of the protected attributes by both evaluators.
We define \textit{agreement} as the proportion of evaluations in which both evaluators have chosen the same value for the protected attribute, for instance if both evaluators considered a specific Nurse image having the gender male we consider they agreed.
We can think of the agreement as an analogous to accuracy in a classification task setting.

In Table \ref{tab:agreement_chi} we present the agreement of both human evaluators (Eval. 1 and Eval 2) and also the agreement of each evaluator with the GPT4o model.

\begin{table*}[htb]
\centering

\caption{Comparison between each human evaluator and the GPT4o model regarding protected attributes. We consider that evaluators agree when they annotate an image with the same protected attribute. Both the percentage of agreement and $\chi^2$ test results are presented. We highlight in \colorbox{greenpastel}{green} comparisons that differences were not statistically significant and in \colorbox{redpastel}{red} differences with statistical significance ($\text{p-value} < .05$).}
\begin{tabular}{@{}lllrrr@{}}
\toprule
\multicolumn{1}{c}{Concept (Bias)}  & \multicolumn{1}{c}{Model} & \multicolumn{1}{c}{Comparison} & \multicolumn{1}{c}{Agreement} & \multicolumn{1}{c}{$\chi^2$} & \multicolumn{1}{c}{p-value} \\ \midrule
\multirow{6}{*}{Nurse (Gender)}     & \multirow{3}{*}{Original} & GPT4o and Eval. 1               & 100.00\%                      & 0.00                     & \cellcolor{greenpastel}{1.000}                       \\
                                    &                           & GPT4o and Eval. 2               & 98.99\%                       & 0.50                     & \cellcolor{greenpastel}{0.478}                       \\
                                    &                           & Eval. 1 and Eval. 2               & 98.99\%                       & 0.50                     & \cellcolor{greenpastel}{0.478}                       \\
                                    & \multirow{3}{*}{Debiased} & GPT4o and Eval. 1               & 96.48\%                       & 0.00                     & \cellcolor{greenpastel}{1.000}                       \\
                                    &                           & GPT4o and Eval. 2               & 90.45\%                       & 3.24                     & \cellcolor{greenpastel}{0.072}                       \\
                                    &                           & Eval. 1 and Eval. 2               & 91.96\%                       & 2.68                     & \cellcolor{greenpastel}{0.102}                       \\ \midrule
\multirow{6}{*}{Firefighter (Race)} & \multirow{3}{*}{Original} & GPT4o and Eval. 1               & 95.98\%                       & 3.70                     & \cellcolor{greenpastel}{0.295}                       \\
                                    &                           & GPT4o and Eval. 2               & 95.98\%                       & 1.44                     & \cellcolor{greenpastel}{0.695}                       \\
                                    &                           & Eval. 1 and Eval. 2               & 95.98\%                       & 1.36                     & \cellcolor{greenpastel}{0.507}                       \\
                                    & \multirow{3}{*}{Debiased} & GPT4o and Eval. 1               & 62.81\%                       & 62.16                    & \cellcolor{redpastel}{2.03e-13}                    \\
                                    &                           & GPT4o and Eval. 2               & 59.30\%                       & 63.42                    & \cellcolor{redpastel}{1.09e-13}                    \\
                                    &                           & Eval. 1 and Eval. 2               & 74.37\%                       & 8.32                     & \cellcolor{redpastel}{0.040}                       \\ \midrule
\multirow{6}{*}{CEO (Age)}          & \multirow{3}{*}{Original} & GPT4o and Eval. 1               & 23.81\%                       & 224.95                   & \cellcolor{redpastel}{1.42e-49}                    \\
                                    &                           & GPT4o and Eval. 2               & 67.20\%                       & 52.71                    & \cellcolor{redpastel}{3.59e-12}                    \\
                                    &                           & Eval. 1 and Eval. 2               & 49.74\%                       & 88.08                    & \cellcolor{redpastel}{7.49e-20}                    \\
                                    & \multirow{3}{*}{Debiased} & GPT4o and Eval. 1               & 79.89\%                       & 27.54                    & \cellcolor{redpastel}{1.04e-06}                    \\
                                    &                           & GPT4o and Eval. 2               & 94.71\%                       & 1.09                     & \cellcolor{greenpastel}{0.581}                       \\
                                    &                           & Eval. 1 and Eval. 2               & 82.54\%                       & 24.26                    & \cellcolor{redpastel}{5.40e-06}                    \\ 
                                    \bottomrule
\end{tabular}
\label{tab:agreement_chi}
\end{table*}

We see that the agreement varies a lot between each of the evaluated concepts, for instance in gender both Eval 1 and Eval 2 agreed in the protected evaluation in more than 90\% of the cases, however in age the agreement was only about 50\% of cases.
This is probably due to the fact that is harder to define age from a single image than it is to define gender among the generated images.

Another interesting trend is that agreement seems to be lower in the debiased model in comparison to the original one.
This happened both in gender and race, but did not happen in the age debiased model.
We suspect that the debiasing procedure created images with a mixture of characteristics of protected attributes and resulted in images harder to evaluate.

We also present the results of $\chi^2$ test in Table \ref{tab:agreement_chi}.
In this setup the null hypothesis is that Eval. 1 and Eval. 2 are equivalent, we reject the null hypothesis if the significance $\text{p-value}<0.05$.
We fail to reject the null hypothesis in the cases where Eval. 1 and Eval. 2 are annotating gender and when they are annotating race for the original model, in the other cases we reject the null hypothesis and there appears to be a statistical significant difference between human annotators when evaluating age.
We also highlight that the p-value for race evaluation of the debiased model was the closest result to our significance threshold of $0.05$.

Our decision to not apply a correction for multiple tests, e.g. Bonferroni directly impacted one line of Table \ref{tab:agreement_chi}, the Eval 1 and Eval 2 agreement on the debiased firefighter.
By not applying Bonferroni, our results were statistically significantly distinct but would not be otherwise.
This difference does not invalidate our results, since all the other cases remain the same.

\subsection{Agreement between Human Evaluators and GPT4o}

Here we compare Eval. 1 and Eval. 2 to the GPT4o evaluation.
By inspecting the Table \ref{tab:agreement_chi} we see that the overall trend is that GPT4o agrees with both evaluators, specially in the protected attribute gender on images of nurses.
The lowest agreement was 23\% with Eval. 1 regarding age. 
Nevertheless, it is worth noting that age was also the protected attribute with lowest agreement between human evaluators (49.74\%), which shows that this was difficult task.

When comparing annotations from Eval. 1 and Eval. 2 to GPT4o, the $\chi^2$ test also evaluates the null hypothesis that GPT4o evaluation is equivalent to a human evaluator.
In images of nurses, there was no significant statistical difference between the evaluation from GPT4o and any of the evaluators. 
The results varied in images of firefighters. In the original model, there was no significant statistical difference, whereas in the debiased model, a statistical difference was observed between both annotators.
For CEO images, we failed to reject the null hypothesis for the GPT4o and Eval. 2 comparison, which indicates that Eval. 2 labeled images more similarly to GPT4o than Eval. 1.

Overall, we highlight that the agreement between Eval. 1 and Eval. 2 was 82.11\% and 82.96\% regarding original and debiased models respectively, whereas the agreement between GPT4o and Eval 2. was 87.73\% and 81.26\%. 
We see that the agreement between Eval. 1 and Eval. 2 was similar to the agreement between Eval. 2 and GPT4o. This is interesting because it provides evidence in favor of evaluating fairness using models such as GPT4o.

These results indicates that using vision language models such as GPT4o as judges or evaluators in scenarios for fairness evaluation can be a viable option, that is not significantly different from human evaluators in many scenarios.
This is an interesting finding because automatic evaluation of fairness criteria facilitate fairness evaluation and bias monitoring.

\subsection{Image Quality Evaluation}

\begin{figure}[htb]
    \centering
    \includegraphics[width=\linewidth]{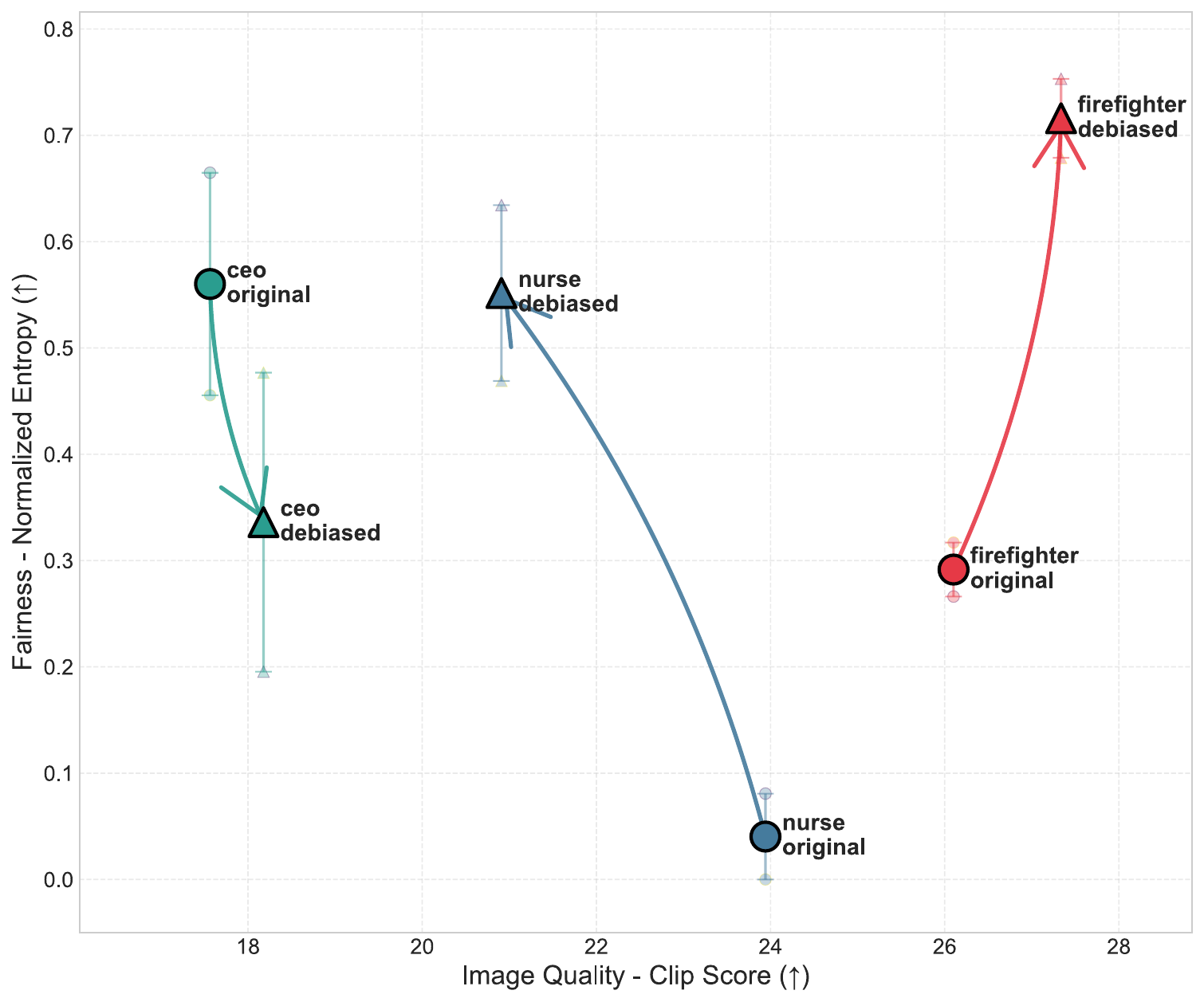}
    \caption{Summary of the debiasing results and image quality. We can see that on two out of three of the concepts evaluated fairness was improved with little to no drop in image quality. Only for the CEO concept the images were equaly biased, but this time toward a distinct valuation of the protected attribute.}
    \label{fig:results-image-quality}
\end{figure}

To evaluate image quality and fairness, we computed both the CLIP-Score and Normalized KL-Divergence for all of the generated images.
Figure \ref{fig:results-image-quality} shows how this metrics behave both for the original model and for the debiased model.

For both the Nurse and Firefighter concepts, our method increased fairness while minimally affecting image quality.
However, in CEO concept even our model being able to visibly change the protected attribute (see Figure \ref{fig:debias_images}), we also generated a biased sample of images, but with a distinct bias. 
This is visible in Figure \ref{fig:results-image-quality}, as the points representing both the original and the debiased models appears with similar fairness score, however the evaluators showed more agreement after the debiasing procedure.

\begin{figure}[htb]
    \centering
    \includegraphics[width=1\linewidth]{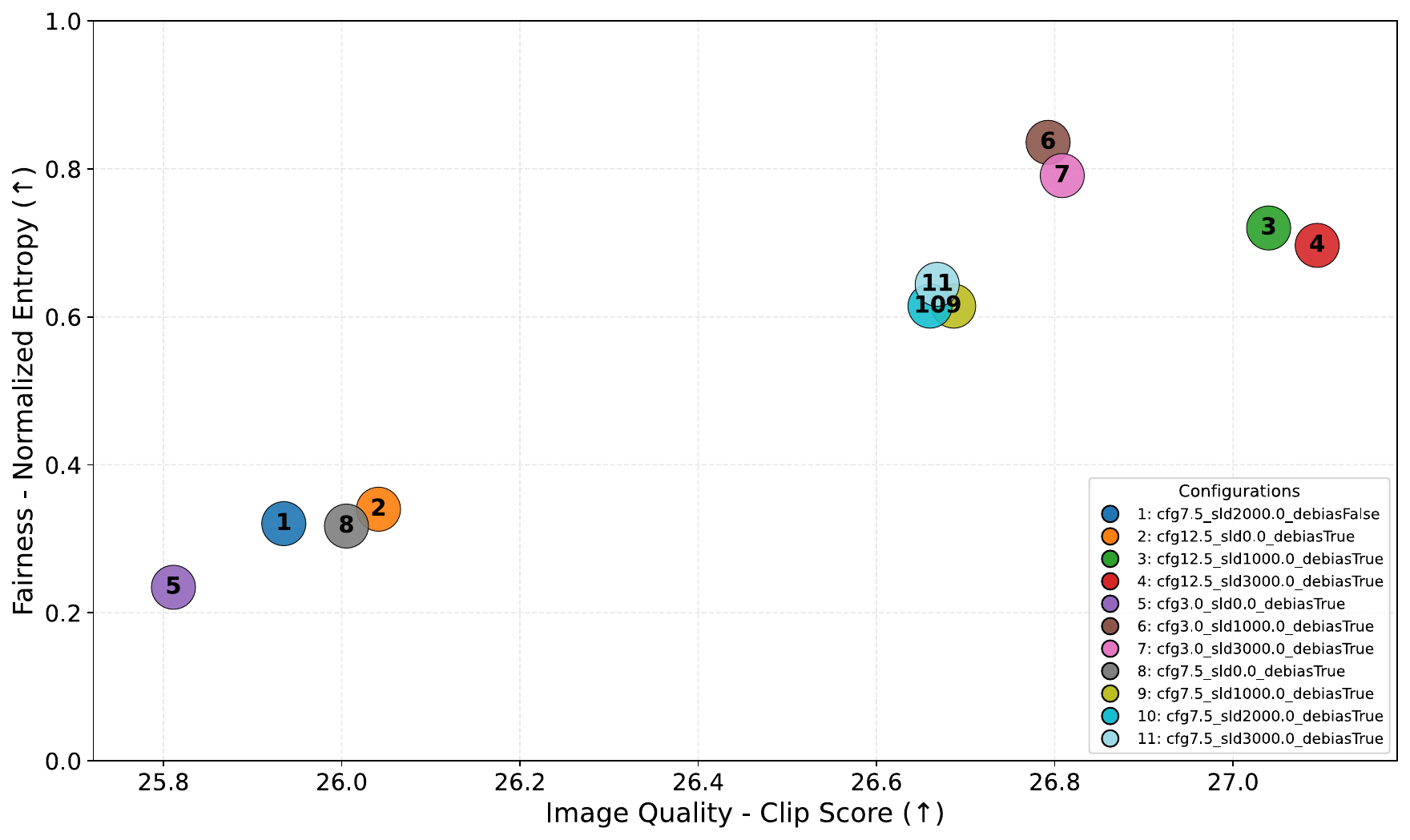}
    \caption{Visualization of the trade-off between image quality and fairness regarding our method. Hyperparametrization two reffers to the original model without debiasing, the other hiperparametrizations refer to the debiased models.}
    \label{fig:many-hyperparameters}
\end{figure}

We also changed the hyperparameters of our debiased procedure and studied how image quality and fairness behaved on distinct hiperparametrization.
Figure \ref{fig:many-hyperparameters} shows a summary of these distinct hiperparametrization.
We see that the original model, marked with the number one, is almost the lowest performing regarding fairness, and several hiperparametrization of the debiasing procedure lead to increased fairness among the generated images

\section{\label{chap:conlcusion}Conclusion}

Fairness in image generation models is a pressing societal concern. 
As generative models become widespread, biases can be propagated and replicated on a large scale.
However, most debiasing methods are prohibitively expensive for practitioners because of high compute demand.
In this study, we presented \textsc{DeCoDi}, an effective debiasing procedure for diffusion-based image generation models that do not require changes in pre-training or fine-tuning. 
It requires only changing the model inference, creating a negligible compute overhead.

Two independent human evaluators assessed our debiasing procedure and agreed that our model generated more diverse images regarding the protected attributes of gender, age, and ethnicity. 
Although the debiasing procedure changed the protected attributes, we could not detect changes in other characteristics of the generated images such as pose, facial expression, and background.

It is important to highlight that despite the good results, there is an inherent subjectivity in fairness and bias evaluation.
This subjectivity was evident by the lack of agreement between human evaluators, especially in the protected attributes of age and race.
Even though the evaluators can disagree, our debiasing method generates more diverse images of the evaluated protected attributes.

Our evaluation comparing GPT4o to human evaluations shows that it is not always possible to differentiate a human evaluation than an evaluation performed by a vision-language model. 
This is interesting because we are providing evidence that this type of bias evaluation using model as a judge can be a viable option for identifying biases in image generation models.

Fairness in generative models is a complex and nuanced subject that is far from solved.
However, \textsc{DeCoDi} is a promising method that can mitigate biases in models without the need for pre-training or fine-tuning.

\begin{credits}
\subsubsection{\ackname} 

This work was funded by Motorola Mobility Brazil. 
This study was financed in part by the Coordenação de Aperfeiçoamento de Pessoal de Nível Superior - Brasil (CAPES) - Finance Code 001.
During the drafting of this article, LLMs were used solely for grammar review, clarity improvement, and spelling correction.

\subsubsection{\discintname}
The authors declare no competing interests.

\end{credits}
%
%
%
\bibliographystyle{splncs04}
\bibliography{bib}

\end{document}